\documentclass[]{article}
\usepackage{amsmath}
\usepackage{amssymb}
\usepackage{amsthm}
\usepackage{graphicx}
\usepackage{float}
\usepackage{multirow}
\usepackage{kbordermatrix}
\usepackage{amsmath}

\title{A Statistical Model for Stroke Outcome Prediction and Treatment Planning}
\author{
Abhishek Sengupta$^{1}$, Vaibhav Rajan$^{1}$, Sakyajit Bhattacharya$^{1}$, GRK Sarma$^{2}$
\thanks{$^{1}$Xerox Research Centre India (XRCI), Bangalore, India
        {\tt\scriptsize \{abhishek.sengupta,vaibhav.rajan,sakyajit.bhattacharya\} at xerox.com}}%
\thanks{$^{2}$St. John's Medical College Hospital, Bangalore, India
        {\tt\footnotesize grksarma at yahoo.com}}%
}
\date{}
\begin{document}
\maketitle

\begin{abstract}
Stroke is a major cause of mortality and long--term disability in the world. Predictive outcome models in stroke are valuable for personalized treatment, rehabilitation planning and in controlled clinical trials. In this paper we design a new model to predict outcome in the short-term, the putative therapeutic window for several treatments. Our regression-based model has a parametric form that is designed to address many challenges common in medical datasets like highly correlated variables and class imbalance. Empirically our model outperforms the best--known previous models in predicting short--term outcomes and in inferring the most effective treatments that improve outcome. 
\end{abstract}

\section{Introduction}\label{intro}

Stroke is the second--leading cause of death and
the leading cause of serious long--term disability in the world;
it is the fifth--leading cause of death in the USA
with an estimated annual economic burden of \$34 billion \cite{mozaffarian2015heart}.
About 87\% of all strokes are ischemic strokes where blood flow to 
the brain is blocked \cite{mozaffarian2015heart}.
Discovering risk factors, 
predicting outcome, mortality and complications,
planning patient rehabilitation and treatment are all active areas of research
within both medical and machine learning communities \cite{khosla2010,letham2013,saran2014}.

Stroke impairs many critical neurological functions, causing a broad range of physical and social disabilities. 
The final outcome after a stroke can range from complete recovery to permanent disability and death. 
Accurate outcome prediction has several uses: to guide treatment decisions,
set prognostic expectations, plan rehabilitation, and select patients in controlled clinical trials.
Many outcome models have been proposed that differ mainly in the risk factors
used as predictors,
for example, \cite{johnston2000predictive,baird2001three,reid2012developing} that use clinical and imaging variables to predict outcome at 3--6 months after stroke,
and \cite{counsell2002predicting,konig2008predicting,muscari2011simple} using only
a few clinical variables for predicting outcome and mortality at 3--9 months.

The modified Rankin scale, shown in table \ref{rankin},
is a quantified measure of disability and has been widely used to 
evaluate stroke outcomes \cite{van1988interobserver}.
The validity and reliability of the scale has been extensively studied
and attested \cite{Banks2007}.

\begin{table}[!h]
\footnotesize
\centering
\begin{tabular}{|p{0.25cm}|p{7cm}|}
\hline
1 & No symptoms at all\\
\hline
2 & No significant disability despite symptoms; able to carry out all usual duties and activities\\
\hline
3 & Slight disability; unable to carry out all previous activities, but able to look after own affairs without assistance\\
\hline
4 & Moderate disability; requiring some help, but able to walk without assistance\\
\hline
5 & Moderately severe disability; unable to walk without assistance and unable to attend to own bodily needs without assistance\\
\hline
6 & Severe disability; bedridden, incontinent and requiring constant nursing care and attention\\
\hline
7 & Dead\\
\hline
\end{tabular}
\caption{\footnotesize Modified Rankin Scale \cite{van1988interobserver}}
\label{rankin}
\end{table}

The aim of this work is to design a statistical model 
that can (1) predict short--term stroke outcome in a patient and
(2) infer the treatments that are most influential in affecting outcome.
Stroke treatment must be tailored to the individual based on identification of the risk of damage and estimation of potential recovery \cite{Farr1998}, and is one of the most important uses of outcome models. 
However existing outcome models use only a small number of predictive factors and are considered to be unreliable for guiding treatments (see \cite{counsell2004predicting,dennis2008predictions} for details).
In this study, we model the outcome using 6 past conditions, 3 demographic variables, 
16 clinical variables, 23 treatment variables and the admission Rankin score (that measures
the patient's initial condition). 

Unlike previous outcome models, our model is designed to learn the factors that affect outcome
in the short--term, between admission and discharge, that is believed to be the
therapeutic window for neuroprotective drugs and thrombolysis
\cite{weimar2004},
although our model can be used to predict long--term effects as well.
To our knowledge, no previous work has studied the combined effects of 
risk factors and 
treatment options, for short--term outcome prediction.




\subsubsection{Modeling Challenges}
The modeling problem can be viewed as follows.

\begin{figure}[h]
\centering
\includegraphics[width=0.45\columnwidth]{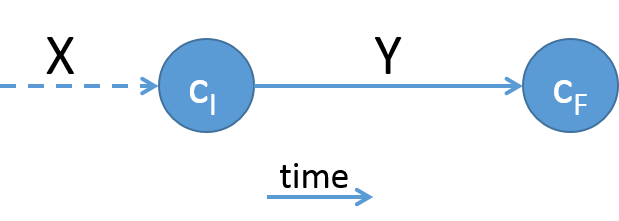}
\end{figure}

A subject with initial condition $c_I$ progresses to a final condition $c_F$. 
During this transition the subject's state and the factors 
affecting it are represented by features $Y$.
The state of the subject before condition $c_I$ is represented by features $X$.
Conditions $c_I,c_F$ are typically quantified by a discrete, ordinal scale.

Such problems occur commonly, though not exclusively, in healthcare.
E.g. in this study, $c_I$ is quantified by the patient's Rankin score at hospital admission, 
and $c_F$ by the Rankin score at discharge, $X$ represents features like
past conditions and $Y$ represents treatments or procedures undertaken
during the hospital stay.
Our aim is to build a predictive model for (the value at) $c_F$ using 
predictors {$X,Y,c_I$}, while addressing the following challenges.
\begin{itemize}
\item
{\it High Correlation.} When the time elapsed between $c_I$ and $c_F$ is short, e.g. between
hospital admission and discharge, and if the effects of $Y$ are not
easily discernible, $c_I$ and $c_F$ are highly correlated. In such cases,
using $c_I$ as a predictor of $c_F$ (like in a simple Logistic Regression model)
results in it masking the effect of all the variables in $Y$.
This results in trivial predictions (predicted value of $c_F$ equals that of input $c_I$)
and makes it impossible to infer the effects of $Y$ on the final condition.

Note there are no assumptions on the correlation among the predictors $X,Y,c_I$
and techniques like Principal Components Analysis (PCA) may be applied to obtain uncorrelated features. 
This still does not solve the problem of high correlation with the outcome variable, $c_F$.
\item
{\it Multiple Classes.}
Each of the states can have multiple levels, e.g. 7 levels for Rankin score.
It is possible to build separate classification models for each pair $(c_I,c_F)$ 
or for each initial condition $c_I$, but such models are unable to utilize
all the information present in the data, as we show empirically.
\item
{\it Data Imbalance.}
Due to the high correlation present, most of the subjects begin and end
with the same level in $c_I, c_F$. Observations with different 
conditions, $c_I \neq c_F$, are far fewer in number leading to the classical
problem of imbalance prevalent in many medical datasets \cite{uguroglu2012cost,reddybook}.
\item
{\it Small Datasets.}
Many clinical studies are conducted on small groups of volunteers and such
datasets are typically small. The presence of imbalance, multiple classes,
correlations and high dimensional features makes the modeling task even more challenging
on such datasets.
\item
{\it Missing Data.}
Medical datasets often contain missing values mainly because not all 
measurements/investigations are done for all patients. 
While many previous works have addressed missing value imputation
for continuous data, few have satisfactorily dealt with discrete data.
\end{itemize}

\subsubsection{Our Contributions}
\begin{enumerate}

\item
We present a new regression--based model for predicting 
the value of $c_F$ using $X,Y,c_I$.
When used for predicting short--term stroke outcome
it achieves significantly higher accuracy than the best--known
previous models.
The parametric form of our model is designed to address the challenges listed 
that are prevalent in clinical datasets.

\item
Our model allows us to infer the most effective treatments in
influencing outcome and are 
independently validated by previous clinical studies.
Other competitive models are unable to produce similar 
clinically justifiable inferences.

\item
We present a new technique for imputing missing categorical data, that achieves higher accuracy than state-of-the-art multiple imputation methods.

\end{enumerate}

\section{Our New Model} \label{prediction model}

Let $N$ be the total number of observations in the (training) data.
Let $X_{N \times L}$ denote the feature matrix relevant prior to condition $c_I$, 
we use $x_k$ to denote the $L$--dimensional feature vector for observation $k$.
Similarly, we use $Y_{N \times M}$ and $y_k$, respectively, to denote the feature matrix and
$M$--dimensional feature vector for observation $k$, 
relevant between $c_I$ and $c_F$.
See table \ref{data} for examples of such features in our stroke dataset.
Let conditions $c_I, c_F$ be measured on an integral scale of 1 to $K$.
The $K \times K$ contingency table, $\Omega$, below shows the number of subjects (in training),
$n_{i,j}$ with condition $c_I = i$ and condition $c_F = j$,
where $i = 1, \ldots K; j = 1, \ldots K$.

\begin{equation*}
\kbordermatrix{
    \mbox{Score}&1 & .. & j & .. & K\\
    1 & n_{1,1}   & .. &  n_{1,j} &  ..  &  n_{1,K}  \\
    : & ..   &  ..  &  ..  & .. &  .. \\
    i & n_{i,1}  &  ..  &  n_{i,j}  & ..  & n_{i,K} \\
    : & .. & .. & .. & .. & .. \\
    K & n_{K,1} &  .. &  n_{K,j}  &  ..  &  n_{K,K}\\     
}
\end{equation*}



Given the data, the likelihood can be modelled as a multinomial probability subject to the condition $ \sum\limits_{j=1}^K p(i \rightarrow j|x_{k} , y_{k}) = 1, \forall i, k.$ 
(ignoring parameter–-independent normalization constants):

{\small $$ L(\theta) \propto \prod\limits_{k=1}^N \prod\limits_{i=1}^K \prod\limits_{j=1}^{K-1} \{{p(i \rightarrow j| x_{k} , y_{k})}^{ r_{k,i} c_{k,j}}\} \{1 - \sum\limits_{j=1}^{K-1}{p(i \rightarrow j| x_{k} , y_{k})}\}^{ r_{k,i} c_{k,K}} $$}



 $ r_{k,i}=\begin{cases}
    1, & \text{if the $k^{th}$ observation has $c_I = i$ ($i^{th}$ row) }\\
    0, & \text{otherwise}
  \end{cases} $
  
  $ c_{k,j}=\begin{cases}
    1, & \text{if the $k^{th}$ observation has $c_F = j$ ($j^{th}$ column)}\\
    0, & \text{otherwise}
  \end{cases} $
  
where $p(i \rightarrow j|x_{k} ,y_{k})$ is the probability of a 
subject with condition $c_F = j$, given features $x_k, y_k$ and initial condition $c_I = i$.
We propose the following form for $p(i \rightarrow j|x_{k} ,y_{k})$:
\begin{equation*}
p(i \rightarrow j| x_{k} ,y_{k})=\frac{\lambda_{i}(x_{k})}{\alpha+\lambda_{i}(x_{k})+ K_{ij} \, \gamma_{j}( y_{k})}
\end{equation*}
\noindent
where:
\begin{flushleft}
$\lambda_{i}({x_{k}}) = \text{exp}(\beta_{0i} + \sum\limits_{l=1}^L \beta_{li}\,x_{kl})$, denotes the
effects of features $x_k$ on the initial condition $c_I$ ($i^{th}$ \emph{row effect}). 
$ \gamma_{j}({y_{k}}) = \text{exp}(\delta_{0j} + \sum\limits_{m=1}^M \delta_{mj}\,y_{km})$, denotes the
effects of features $y_k$ on the final condition $c_F$ ($j^{th}$ \emph{column effect}).\\ 
$ K_{ij} = C \* \{(j-i)^{2} +1\} $;
C and $\alpha$ are constants.
\end{flushleft}

We now justify our choice of this parametric form.

\begin{enumerate}

\item The row effects $\lambda_{i}(x_{k})$ depend only on features relevant before $c_I$ while the column effects
 $ \gamma_{j}(y_{k})$ depend on features relevant between $c_I$ and $c_F$.
\item One of our objectives is to also find which features $y_k$ lead to an improvement in the subject's condition, i.e. $c_F < c_I$. Since $y_k$ appears only in the denominator, feature coefficients with the least values (most negative) will have the maximum impact in improvement.
\item
Since the model is purely probabilistic, distributions of these coefficients can be derived 
and hence p--values can be computed, to test the significance of the features.
\item 
In small datasets with imbalance, the contingency matrix has higher numbers along the diagonals
(i.e., most subjects with $c_I = i$ remain at $c_F = i$), with the number of subjects decreasing gradually as we move away from the diagonal (e.g. a transition from 3, to 2 or 4 is more likely than to 1 or 5). 
Observing this pattern, we deliberately introduce the term $K_{i,j}$, which increases the probability value if \textit{i} is close to \textit{j}.
The constant $C$ plays the role of weights used in cost--sensitive learning algorithms 
for imbalanced data classification \cite{zadrozny2003cost}. 
In our case the imbalance is between diagonal and off--diagonal elements in the contingency table. 
A higher value of $C$ increases the weight of off--diagonal elements which helps when there are very few 
off--diagonal elements.
Strategies similar to those used in imbalanced data learning can be adopted to choose C, e.g. 
C can be chosen to be $\frac{n_d}{n_o}$ where $n_d = \Sigma n_{i,i}$ is the number of diagonal elements and 
$n_o = N - n_d$, is the number of off--diagonal elements.
\item The constant $\alpha = 0.001$ is added to the denominator to handle model identifiablity problems. 
Without the term, multiplying the numerator and denominator by a constant does not affect the probability but changes the coefficient estimates, necessitating constraints on the parameters. Thus, we add a small constant that ensures unique estimates without affecting the likelihood.
\end{enumerate}

\subsubsection{Advantages of our Approach}

\begin{itemize}
\item
By not taking $c_I$ as a feature directly, we overcome the problem of $c_I$ being highly correlated to $c_F$.
If $c_I$ were to be taken as a feature in a classifier (like Logistic Regression), it is given the  
the highest significance and the rest of the features are ignored (with zero or nearly zero coefficients). 
Such a classifier fails to predict well for those subjects whose difference in condition ($c_F - c_I$) is non--zero.
Moreover, it fails to infer the effects of $y_k$ on $c_F$.
\item
Simple predictors that predict differences in outcome ($c_F - c_I$) using features $X, Y$ do not take into account the differences between varying initial and final conditions.
For example, a subject with initial condition $c_I = 3$ and final condition $c_F = 1$, is different from a subject 
with $c_I = 4, c_F = 2$ and this can affect the predictive performance as seen in our experiments.
\item
Multiple classifiers can be trained, one for each $c_I$ (row). We find that this approach is severely affected by class imbalance and gives trivial predictions $c_F = c_I$ in almost all cases. In contrast, our approach, for a particular row, takes into account \textbf{all} the observations in the corresponding columns to make the prediction, thus tackling the imbalance and giving better accuracy, as shown empirically. 
This is particularly helpful while learning from small datasets.
\item
To compute the probability, say $p(c_I=2 \rightarrow c_F=1)$, we learn from \textbf{all} the subjects with $c_F = 1$ and from \textbf{all} the subjects with $c_I = 2$, as opposed to learning from only those subjects with $c_F = 2$ and $c_I = 1$ (which a row-wise classifier will do). Intuitively, all subjects with $c_F = 1$ have some distinguishing characteristics (irrespective of their $c_I$ values) and similarly for all subjects with $c_I = 2$, and our model attempts to capture this information.

\end{itemize}

\subsubsection{Estimation}

Maximum likelihood (ML) estimates of parameters are found using gradient ascent. We use elastic net regularization(Ref) with regularization constants $\lambda_{11},\lambda_{12},\lambda_{21}$ and $\lambda_{22}$ ($\lambda_{11}$ and $\lambda_{12}$ are constraints for $\beta$ 's while $\lambda_{21}$ and $\lambda_{22}$ are constraints for $\delta$ 's). 
The log likelihood of the data including the regularization terms is:

{\footnotesize  $$ l(\theta)= \sum\limits_{k=1}^N  \sum\limits_{i=1}^K \sum\limits_{j=1}^{K-1} r_{k,i} c_{k,j} \Bigg[ \beta_{0i} + \sum\limits_{l=1}^L \beta_{li}\,x_{kl} - \mathrm{log}\{\mathrm{exp}(\beta_{0i} + \sum\limits_{l=1}^L \beta_{li}\,x_{kl})$$\\
$$+ K_{ij} \, \mathrm{exp}(\delta_{0j} + \sum\limits_{m=1}^M \delta_{mj}\,y_{km}) +\alpha\}\Bigg] \ +  \sum\limits_{k=1}^N  \sum\limits_{i=1}^K  r_{k,i} c_{k,K} $$
$$log \lbrace 1 - \sum\limits_{j=1}^{K-1} \frac{\text{exp}(\beta_{0i} + \sum\limits_{l=1}^L \beta_{li}\,x_{kl})}{\alpha+\text{exp}(\beta_{0i} + \sum\limits_{l=1}^L \beta_{li}\,x_{kl})+ K_{ij} \, \text{exp}(\delta_{0j} + \sum\limits_{m=1}^M \delta_{mj}\,y_{km})} \} $$
$$ + \lambda_{11} \sum\limits_{i=1}^K\sum\limits_{l=1}^L |{\beta_{li}}| + \lambda_{12} \sum\limits_{i=1}^K\sum\limits_{l=1}^L {\beta_{li}}^{2} +  \lambda_{21} \sum\limits_{j=1}^{K-1}\sum\limits_{m=1}^M |{\delta_{mj}}| + $$
$$\lambda_{22}\sum\limits_{j=1}^{K-1} \sum\limits_{m=1}^M {\delta_{mj}}^{2} $$
}

ML estimates have to be obtained for the
following $K(L+M+2)-(M+1)$
parameters $\theta=\{\beta_{0i},\beta_{li},\delta_{0j},\delta_{mj}\},
i = 1,\ldots,K; j = 1, \ldots ,K-1; l = 1,\ldots,L; m = 1,\ldots,M$
(since the probabilities sum to 1 in each row, $j$ iterates over only $K-1$ entries). 
We equate the partial derivatives to zero but these partial derivatives do not have a closed form solution, so we use gradient ascent to solve for the parameters iteratively. The regularization constants are chosen empirically.
The complete derivation is shown in the appendix.
%
%

\subsubsection{Computational Complexity}
For a $K \times K$ contingency table, $N$ observations and feature matrices
$X_{N \times L}$ and $Y_{N \times M}$, the complexity of
estimating all the parameters of our model is $\mathcal{O}(t K^2 N (L + M))$
where $t$ is the number of iterations of the gradient ascent.

\subsubsection{Outcome Prediction and Feature Importance}
To predict the final condition $c_F$ of a subject given $c_I, x_k, y_k$,
we compute $\hat{p}(i \rightarrow j|x_{k} ,y_{k})$ from our model 
(using ML estimates of coefficients), for all $K$ values of \textit{j}.
The predicted final condition is the value $j$ with the maximum probability.

To evaluate the features that lead to improvement in a subject's condition,
instead of using the final condition $c_F$, we use 
an outcome variable $\Delta$,
set as follows: $\Delta = 1$ if $c_F - c_I > 0$,  
$\Delta = -1$ if $c_F - c_I < 0$ 
and $\Delta = 0$ if $c_F = c_I$.
The contingency table and model coefficients are recomputed as before.
The interpretation of ${p}(i \rightarrow \Delta|x_{k} ,y_{k})$ now changes to the 
probability of a subject's condition to change \textit{by} $\Delta$
(negative indicating improvement, positive indicating deterioration
for the Rankin scale) 
given the initial condition \textit{i} and features $X,Y$. 
To find which features lead to an improvement in outcome, i.e. $\Delta = -1$, 
we select the features with
the smallest $\delta_{m \Delta}$ (coefficients of $y_{km}$) values, with $\Delta = -1$.
Influence of feature interactions in the model 
can be studied by adding additional variables (e.g. $y_{1k}* y_{2k}$) 
in the {y}$_{k}$ vector. 

\subsubsection{P value Computation}
In order to test the significance of a parameter $\delta_{mj},\,\, (m = 1,\ldots,M; j = 1,\ldots,K)$ pertaining to 
the features $Y$, we need to test the null hypothesis 
$H_{0}: \delta_{mj} = 0 $ against the alternative hypothesis $H_{1}: \delta_{mj} \neq 0$. 
Let $\widehat{\delta_{mj}}$ denote the observed value and let $T = \widehat{\boldsymbol{\delta_{mj}}}$, 
the ML estimator, denote our test statistic.
The p-value for the two--sided test is  $2 \times \text{min} \lbrace P_{H_0} ( T > \widehat{\delta_{mj}}) ,P_{H_0} ( T <  \widehat{\delta_{mj}}) \rbrace $.

Since the null distribution is not known and we do not have a closed form 
expression for $T$, we use bootstrapping to compute an empirical p--value.
We simulate multiple contingency tables $\Omega_b, \, b =1,\ldots,500$, and
estimate the distribution of $T$ from $\Omega_b$, under the null hypothesis
$\delta_{mj} = 0$, without changing the features $X,Y$.
Each table $\Omega_b$ is initialized to all zeros and 
then selected table entries $(i,j)$ ($i^{th}$ row, $j^{th}$ column) are incremented
by 1 in the following way.
For each observation with initial condition $c_I = i$,
we generate one sample from a multinomial distribution, 
with probability $\hat{p}(i \rightarrow j|x_{k},y_{k},\delta_{mj} = 0))$,
and obtain a value of $c_F = j$, which gives the selection $(i,j)$. 
In total $N$ incrementations are done, once for 
each observation in our real dataset.
We thus obtain a bootstrap contingency table $\Omega_b$ and
different final conditions $c_F$ on which we train our model (retaining
all other inputs as given)
to obtain $\widehat{\delta_{mj}^b}$, for the $b^{th}$ bootstrap, which is a sample from $T$ 
under the null hypothesis. 
The empirical p-value is 
$2 \times min\{ (\sum\limits_{b=1}^{500}  {\bf I}_{T > \widehat{\delta_{mj}^b}})/500,(\sum\limits_{b=1}^{500}  {\bf I}_{T < \widehat{\delta_{mj}^b}})/500\}$;
${\bf I}_c = 1$ if the condition $c$ is true, otherwise 0.

\subsection{Imputation using Association Score}

Let {\bf $(x_{i},z_{i})$} be the $i^{th}$ observation with $z_{i}$ denoting the response variable (the variable to be imputed for feature $Z$), and {\bf $x_{i}$}, the $p$--dimensional vector of predictors.
As in MICE, we impute $z_{i}$ by sampling from one of the values in $Z_{-i}=\{z_{1},\ldots,z_{i-1},z_{i+1},\ldots,z_{N}\}$, the subset of the feature values 
with no missing values, where $N$ is the total number of observations.
For the $k^{th}$ observation with $z_{k}$ not missing, we define a measure of association 
$ \textbf Q_{ik} = \frac{\textbf C_{ik}-\textbf D_{ik}}{\textbf C_{ik}+\textbf D_{ik}} $
where, $\textbf C_{ik} = \sum\limits_{i=1}^p {\bf I}_{x_{k,j}=x_{i,j}}$, $\textbf D_{ik} = \sum\limits_{i=1}^p {\bf I}_{x_{k,j} \neq x_{i,j}}$;
$x_{i,j}$ denotes the $i^{th}$ observation for the $j^{th}$ feature,
the sum being over all \textit{j} for which $x_{k,j}$ and $x_{i,j}$ are not missing
and ${\bf I}_c$ is an indicator function yielding 1 if the condition $c$ is true, otherwise 0.
Thus, $C_{ik}$ and $D_{ik}$ respectively denote the number of concordant and discordant pairs between the $i^{th}$ and the $k^{th}$ observations. $Q_{ik}$ is similar to the Kendall's $\tau$ measure of association, with a higher value of $Q_{ik}$ indicating stronger association. 
Let $Z = \{z_q\}$ be the list of values with $m$ highest association scores $Q_{iq}$.	
To impute the missing value $z_i$, we select a value at random from the list of $Z$.
Imputing a single value takes $\mathcal{O}(Np)$ time.

%
%

%

%
%
%

\section{Stroke Data Analysis}\label{study}

We retrospectively analyze the data of 275 ischemic stroke patients 
in the age group of 45 -- 75,
admitted to St. John's Hospital 
The data includes the variables listed in table \ref{data} for each patient. 
Summary statistics for each of the variables are shown in the appendix.
Numerical attributes like investigations are measured in standard units.
Other attributes are suitably encoded as binary or categorical data.
For example, addictions, preconditions and treatments are binary variables indicating presence or absence. 
Radiology investigations are encoded into categories indicating normal or abnormal results.
Complete details on the encoding are shown in the appendix.
Not all investigations are conducted for all the patients and
the treatment variables differ across patients resulting in a large number of missing values.

For our dataset, the contingency table, $\Omega$ is as follows:

 $$
\kbordermatrix{
    \mbox{Score}&1 &2 &3 &4 &5\\
    1 &22 &0 &0 &0 &0  \\
    2 &40 &29 &1 &0 &0 \\
    3 &1  &44 &25 &0 &2 \\
    4 &0 &0 &38 &34 &2 \\
    5 &0 &0 &2 &15 &20   
}$$


\begin{table}[h]
\footnotesize
\centering
\begin{tabular}{|p{0.75cm}|p{7cm}|}
\hline
\multirow{4}{*}{$X_{N \times L}$} & {\bf Demographic:} Age, Gender, Religion \\
& {\bf Time}: time between stroke event and treatment start\\
& {\bf Addictions:} Smoking, Alcohol\\
& {\bf Preconditions:} Hypertension, Diabetes, Ischemic Heart Condition, Preceding Fever\\
\hline
$c_I$ & {\bf Initial condition:} Rankin Score at admission\\
\hline
\multirow{4}{*}{$Y_{N \times M}$} & {\bf Investigations:}	Hemoglobin,	Total Counts,	Differential Counts, Platelet, Creatinine, Serum Sodium, Cholesterol, High Density Lipoprotein (HDL), Low Density Lipoprotein (LDL), Triglycerides,
Admission Blood Pressure, Electrocardiogram (ECG), Echo, Doppler \\
& {\bf Treatment:} Aspirin, Clopidogrel, Atorvastatin, Finofibrate, Edaravon, Citicoline, Heparin, Dalteparin, Enoxaparine, Warfarin, Acitrom, t-PA, , Dabigatran, Anti-hypertensives (ACEI, Beta Channel Blockers, Diuretics, ARB), Other drugs (Piracetam, Mannitol, B-Complex, Pantoprazole, Antibiotics), Physiotherapy\\
& {\bf Others:} Type of ward, number of days in hospital, Complication\\
\hline
$c_F$ & {\bf Final Condition:} Rankin Score at discharge\\
\hline	
\end{tabular}
\caption{\footnotesize Stroke Dataset Variables.}
\label{data}
\end{table}

\subsubsection{Data Preprocessing}\label{preprocess}
Features with values in less than 25\% of patients are omitted.
We use MICE \cite{buuren2011mice} to impute all continuous valued features and our association based method for imputing 
categorical features.

We group the Rankin scores into 3 groups: $g_1 = \{1\}$, $g_2 =\{2,3\}$ and $g_3 = \{4,5,6,7\}$ and use these as $K=3$ levels. The reduced contingency table is:

$$
\kbordermatrix{
    \mbox{Score}&g_1 &g_2 &g_3\\
    g_1&22 &0 &0  \\
    g_2&41 &99 &2 \\
    g_3&0 &40 &71  
}$$

\subsubsection{Simulated Data}

We generate additional synthetic data to evaluate the classifiers on datasets by varying the total observations $N$ and number of classes (levels in $c_I,c_F$), $K$.
Datasets with different $N$ are obtained by multiplying the reduced contingency table above by 2, 10, 15 and 20.
Datasets with different $K$ are obtained from the dataset with $N = 5500$ by splitting the rows 
while maintaining the pattern
found in the real data -- high diagonal values and off--diagonal values decreasing
with increasing distance from the diagonal.
Features $X$ and $Y$ are generated by sampling from 
6--dimensional normal distributions $\mathcal{N}(\mu^X,I)$ and 
$\mathcal{N}(\mu^Y,I)$ where $\mu^X_i = (i-1 + c_I)$
and $\mu^Y_i = (i+3 - c_F), i = 1,\ldots,6$; $I$ denotes the identity matrix.

\section{Experimental Results} \label{Results}

\subsection{Accuracy of Imputation Method}

To test the performance of our imputation method, 
we select only those 211 observations having no missing values in the (binary) treatment features 
and remove 10\% of the values randomly (using these as missing values) from each feature in $X,Y$, and apply both MICE and our imputation method. 
We check the accuracy of the imputed values which is the proportion of the observations for which the values are correctly imputed. We repeat the experiment 5 times, each time with a different random selection of values to be imputed.
Table \ref{imp} shows the accuracy obtained by MICE, K-Nearest Neighbor (KNN) imputation \cite{torgo2010data}
and our method which significantly outperforms both the methods. 

\begin{table}[h]
\scriptsize
\centering
\begin{tabular}{ccc}
\hline
Method & Mean & SD \\
\hline
KNN & 54.5\% & 5.16 \\
MICE & 59.4\% & 4.02 \\
Our Method & {\bf 68.7\%} & 3.86 \\
\hline
\end{tabular}
\caption{\footnotesize Mean imputation accuracy and standard deviation (SD) over 5 runs of our experiment.}
\label{imp}
\end{table}

We measure the predictive accuracy of our model and compare it with 
baseline models that have been used in previous stroke outcome studies.

\subsubsection{Evaluation Metrics} 
Accuracy is defined as the proportion of test observations correctly classified.
\emph{Overestimate error}  is defined as the proportion of the 
misclassified test samples, out of the total number of samples, for which the predicted 
outcome, $\hat{j}$ is greater than the true outcome, $j$; i.e. $\hat{j} > j$. 
Similarly the \emph{underestimate error} measures the proportion of underestimates, $\hat{j} < j$.
All results shown are over five--fold cross validation.

\subsubsection{Baselines}
As baselines we use classification methods Logistic Regression and Support Vector Machines
in two different ways. First we concatenate $X,Y$, use them as features to predict $(c_F - c_I)$;
we denote these classifiers by $LR$ and $SVM$ respectively.
Next we train 5 classifiers, one for each value of $c_I$ (each row in the contingency table)
and use concatenated features $X,Y$ to predict $c_F$.
Given a test case, based on the admission score ($c_I$), 
we predict using the corresponding classifier.
We denote these classifiers by $LR_{row}$ and $SVM_{row}$ respectively.
Our method is denoted by NEW.
%

\subsubsection{Simulated Data}

\begin{figure}
\centering
\includegraphics[width=0.7\columnwidth]{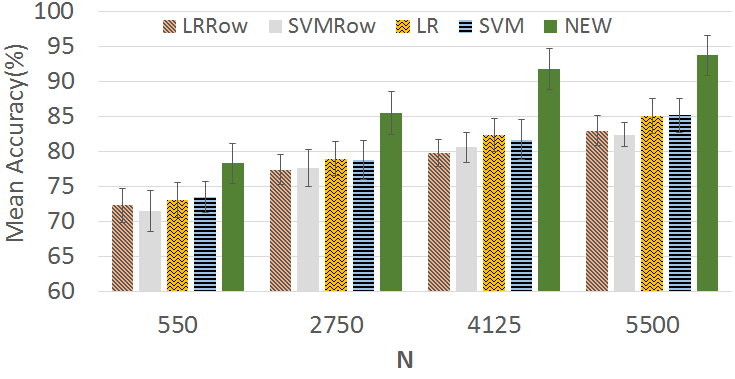}
\includegraphics[width=0.7\columnwidth]{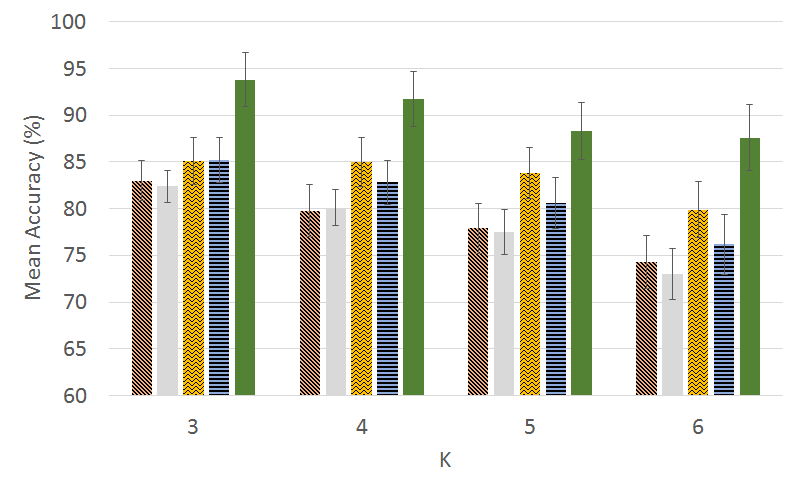}
\caption{\footnotesize Performance on simulated data with fixed K = 3 and increasing N (above) and fixed N = 5500 and increasing K (below).}
\label{fig:simacc}
\end{figure}

The performance of all the classifiers on the simulations 
are shown in figure \ref{fig:simacc}. We see that
the accuracy of NEW is significantly better than
all other baselines and, as expected, the performance improves with more data (increasing $N$).
We also see that the performance of all classifiers deteriorate
as $K$, the number of levels $c_F, c_I$, increase, making multi--class classification harder.
However, our method maintains its superiority over baselines in all four cases.

\subsubsection{Stroke Data}


\begin{figure}
\centering
\includegraphics[width=0.5\columnwidth]{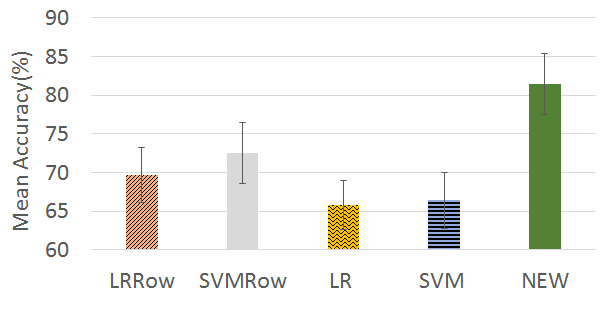}
\includegraphics[width=0.7\columnwidth]{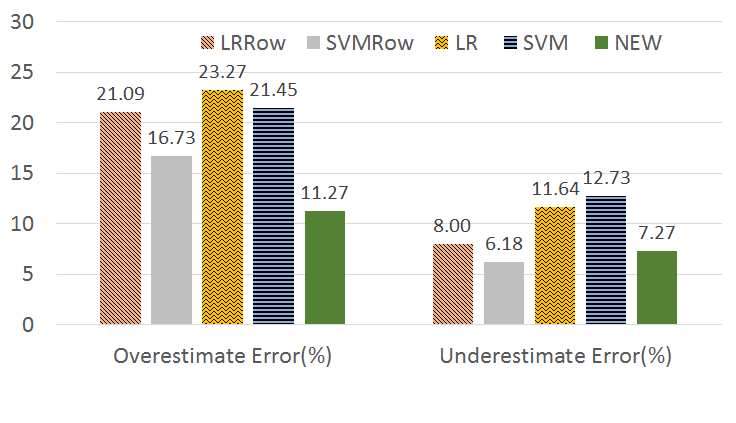}
\caption{\footnotesize Performance on stroke datset: Mean accuracy over 5--fold cross validation (above);
Overestimate and underestimate errors (below).}
\label{fig:realacc}
\end{figure}




Figure \ref{fig:realacc} (above) shows the average predictive accuracy of all the classifiers tested.
Logistic Regression (LR) is the best previously used model for predicting outcome 
 and our model
outperforms LR by nearly 19\%.

Figure \ref{fig:realacc} (below) shows the overestimate and underestimate errors of all the classifiers.
Predicting $\hat{j} > j$ results in overestimating the predicted outcome and may result in overburdening the patients
with unnecessary treatments or higher dosages. Such false predictions are the least ($9.82\%$) in our method.
Underestimating the outcome, $\hat{j} < j$,  may result in undermining the predicted severity of the patient at discharge (or later) and may result in inadequate care. Our model and $SVM_{row}$ have the least number of such false predictions ($< 7\%$).
%

\subsubsection{Analysis of Treatment Effects}

We illustrate the use of our model in analyzing the effects of treatment in stroke outcome.
We train the model as described earlier 
that computes ${p}(i \rightarrow \Delta|\textbf x_{k} , \textbf y_{k})$
where $\Delta$ indicating improvement or deterioration
is used as the outcome variable.

%
%
\begin{table}[h]
\centering
\scriptsize
\begin{tabular}{|c|c|c|}
\hline
Treatment & Coefficient & P--Value \\
\hline
Piracetam & -4.08 & 0.002\\
\hline
Strocit & -3.89 & 0.007\\
\hline
Physiotherapy & -3.61 & 0.009\\
\hline
Fragmin & -3.14 & 0.013\\
\hline
Warfarin & - 2.29 & 0.026\\
\hline
Clexane & -1.67 & 0.031\\
\hline
Acitrom & -1.19 & 0.037\\
\hline
Heparin & -0.76 & 0.045\\
\hline
\end{tabular}
\caption{\footnotesize Treatments with highest impact in improving outcome.}
\label{treatment}
\end{table}

Table \ref{treatment} shows the treatment variables with 
the smallest coefficients in our model along with their p--values.
These treatments have been independently shown to be effective in improving stroke 
outcome in other studies \cite{broderick2002treatment,adams2007guidelines,ginsberg2008neuroprotection,smith2010all}, which provides additional validation for our model.

The Logistic Regression model (LR) is the only other baseline that can be used to infer treatment effects.
It is also the model that has been used in previous stroke outcome studies. 
LR gives Edaravone and Pantoprazole as the only significant factors  with coefficients (pvalues) 0.36 (0.05) and 0.42 (0.007). 
While Edaravone has been found to improve outcome, Pantoprazole is 
given to inhibit gastric acid secretion, and is unrelated to stroke outcome.
Thus LR is unable to identify all the significant factors and also erroneously 
shows an unrelated treatment as significant.

\section{Conclusion}
We develop a new model for predicting stroke outcome that addresses
several challenges common in medical datasets like class imbalance and
highly correlated variables and also design a new imputation strategy
for discrete data. 
Our model is found to be more effective than the best--known previous models 
in predicting short--term outcome and
inferring clinically justified treatment effects.
%
%

\bibliography{strokebib}
\bibliographystyle{plain}

\appendix
\section{Maximum Likelihood Parameter Estimation}
The log likelihood of the data including the regularization terms is given by:

{\footnotesize  $$ l(\theta)= \sum\limits_{k=1}^N  \sum\limits_{i=1}^K \sum\limits_{j=1}^{K-1} r_{k,i} c_{k,j} \Bigg[ \beta_{0i} + \sum\limits_{l=1}^L \beta_{li}\,x_{kl} - \mathrm{log}\{\mathrm{exp}(\beta_{0i} + \sum\limits_{l=1}^L \beta_{li}\,x_{kl}) + K_{ij} \, \mathrm{exp}(\delta_{0j} + \sum\limits_{m=1}^M \delta_{mj}\,y_{km}) +\alpha\}\Bigg] \ + $$
$$ \sum\limits_{k=1}^N  \sum\limits_{i=1}^K  r_{k,i} c_{k,K} log \lbrace 1 - \sum\limits_{j=1}^{K-1} \frac{\text{exp}(\beta_{0i} + \sum\limits_{l=1}^L \beta_{li}\,x_{kl})}{\alpha+\text{exp}(\beta_{0i} + \sum\limits_{l=1}^L \beta_{li}\,x_{kl})+ K_{ij} \, \text{exp}(\delta_{0j} + \sum\limits_{m=1}^M \delta_{mj}\,y_{km})} \} $$

$$ + \lambda_{11} \sum\limits_{i=1}^K\sum\limits_{l=1}^L |{\beta_{li}}| + \lambda_{12} \sum\limits_{i=1}^K\sum\limits_{l=1}^L {\beta_{li}}^{2} +  \lambda_{21} \sum\limits_{j=1}^{K-1}\sum\limits_{m=1}^M |{\delta_{mj}}| + $$
$$\lambda_{22}\sum\limits_{j=1}^{K-1} \sum\limits_{m=1}^M {\delta_{mj}}^{2} $$
}

In order to estimate the ML estimates of the parameters $\theta=(\beta_{0i},\beta_{li},\delta_{0j},\delta_{mj})$, we need to equate the following partial derivatives to zero:

{\footnotesize $$ \dfrac{\partial l(\theta)}{\partial \beta_{0i}} = \sum\limits_{k=1}^N \sum\limits_{j=1}^{K-1} r_{k,i} \Bigg[1-\dfrac{\mathrm{exp}(\beta_{0i} + \sum\limits_{l=1}^L {\beta_{li}\,x_{kl})}}{\mathrm{exp}(\beta_{0i} + \sum\limits_{l=1}^L {\beta_{li}\,x_{kl}})+ K_{ij} \, \mathrm{exp}(\delta_{0j} + \sum\limits_{m=1}^M {\delta_{mj}\,y_{km}})+\alpha}\Bigg] $$

$$ \Bigg[ c_{k,j} - c_{k,K} \dfrac{1}{ 1 - \sum\limits_{j=1}^{K-1} \frac{\text{exp}(\beta_{0i} + \sum\limits_{l=1}^L \beta_{li}\,x_{kl})}{\alpha+\text{exp}(\beta_{0i} + \sum\limits_{l=1}^L \beta_{li}\,x_{kl})+ K_{ij} \, \text{exp}(\delta_{0j} + \sum\limits_{m=1}^M \delta_{mj}\,y_{km})}} \times   \frac{\text{exp}(\beta_{0i} + \sum\limits_{l=1}^L \beta_{li}\,x_{kl})}{\alpha+\text{exp}(\beta_{0i} + \sum\limits_{l=1}^L \beta_{li}\,x_{kl})+ K_{ij} \, \text{exp}(\delta_{0j} + \sum\limits_{m=1}^M \delta_{mj}\,y_{km})} \Bigg] $$

{\footnotesize $$\dfrac{\partial l(\theta)}{\partial \beta_{li}} =  \sum\limits_{k=1}^N \sum\limits_{j=1}^{K-1}{r_{k,i} \,x_{kl}\Bigg[1-\dfrac{\mathrm{exp}(\beta_{0i} + \sum\limits_{l=1}^L {\beta_{li}\,x_{kl})}}{\mathrm{exp}(\beta_{0i} + \sum\limits_{l=1}^L {\beta_{li}\,x_{kl}})+ K_{ij} \, \mathrm{exp}(\delta_{0j} + \sum\limits_{m=1}^M {\delta_{mj}\,y_{km}})+\alpha}\Bigg]} $$

$$ \Bigg[ c_{k,j} -  c_{k,K} \dfrac{1}{ 1 - \sum\limits_{j=1}^{K-1} \frac{\text{exp}(\beta_{0i} + \sum\limits_{l=1}^L \beta_{li}\,x_{kl})}{\alpha+\text{exp}(\beta_{0i} + \sum\limits_{l=1}^L \beta_{li}\,x_{kl})+ K_{ij} \, \text{exp}(\delta_{0j} + \sum\limits_{m=1}^M \delta_{mj}\,y_{km})}} \times  \frac{\text{exp}(\beta_{0i} + \sum\limits_{l=1}^L \beta_{li}\,x_{kl})}{\alpha+\text{exp}(\beta_{0i} + \sum\limits_{l=1}^L \beta_{li}\,x_{kl})+ K_{ij} \, \text{exp}(\delta_{0j} + \sum\limits_{m=1}^M \delta_{mj}\,y_{km})} \Bigg] $$
$$ + \lambda_{11} sign(\beta_{li}) + 2 \lambda_{12} \beta_{li} $$

}

{\footnotesize $$ \dfrac{\partial l(\theta)}{\partial \gamma_{0j}} =  \sum\limits_{k=1}^N \sum\limits_{i=1}^{K}{r_{k,i} \Bigg[\dfrac{K_{ij} \, \mathrm{exp}(\delta_{0j} + \sum\limits_{m=1}^M {\delta_{mj}\,y_{km}})}{\mathrm{exp}(\beta_{0i} + \sum\limits_{l=1}^L {\beta_{li}\,x_{kl}})+ K_{ij} \, \mathrm{exp}(\delta_{0j} + \sum\limits_{m=1}^M {\delta_{mj}\,y_{km}})+\alpha}\Bigg]} $$

$$   \Bigg[ c_{k,K} \dfrac{1}{ 1 - \sum\limits_{j=1}^{K-1} \frac{\text{exp}(\beta_{0i} + \sum\limits_{l=1}^L \beta_{li}\,x_{kl})}{\alpha+\text{exp}(\beta_{0i} + \sum\limits_{l=1}^L \beta_{li}\,x_{kl})+ K_{ij} \, \text{exp}(\delta_{0j} + \sum\limits_{m=1}^M \delta_{mj}\,y_{km})}} \times 
  \frac{\text{exp}(\beta_{0i} + \sum\limits_{l=1}^L \beta_{li}\,x_{kl})}{\alpha+\text{exp}(\beta_{0i} + \sum\limits_{l=1}^L \beta_{li}\,x_{kl})+ K_{ij} \, \text{exp}(\delta_{0j} + \sum\limits_{m=1}^M \delta_{mj}\,y_{km})} \ - \ c_{k,j} \Bigg] $$

}

{\footnotesize $$ \dfrac{\partial l(\theta)}{\partial \gamma_{mj}} =  \sum\limits_{k=1}^N \sum\limits_{i=1}^{K}{r_{k,i} \Bigg[\dfrac{y_{km}\,K_{ij} \, \mathrm{exp}(\delta_{0j} + \sum\limits_{m=1}^M {\delta_{mj}\,y_{km}})}{\mathrm{exp}(\beta_{0i} + \sum\limits_{l=1}^L {\beta_{li}\,x_{kl}})+ K_{ij} \, \mathrm{exp}(\delta_{0j} + \sum\limits_{m=1}^M {\delta_{mj}\,y_{km}})+\alpha}\Bigg]}$$

$$	\Bigg[ c_{k,K} \dfrac{1}{ 1 - \sum\limits_{j=1}^{K-1} \frac{\text{exp}(\beta_{0i} + \sum\limits_{l=1}^L \beta_{li}\,x_{kl})}{\alpha+\text{exp}(\beta_{0i} + \sum\limits_{l=1}^L \beta_{li}\,x_{kl})+ K_{ij} \, \text{exp}(\delta_{0j} + \sum\limits_{m=1}^M \delta_{mj}\,y_{km})}} \times $$
 $$ \frac{\text{exp}(\beta_{0i} + \sum\limits_{l=1}^L \beta_{li}\,x_{kl})}{\alpha+\text{exp}(\beta_{0i} + \sum\limits_{l=1}^L \beta_{li}\,x_{kl})+ K_{ij} \, \text{exp}(\delta_{0j} + \sum\limits_{m=1}^M \delta_{mj}\,y_{km})} \ - \ c_{k,j} \Bigg] $$
 
 $$ +  \lambda_{21} sign(\delta_{mj}) + 2 \lambda_{22} \delta_{mj} $$
}

These partial derivatives do not have a closed form solution, so we use the gradient ascent method to solve for the parameters iteratively, the recursion relation at the $t^{th}$ iteration given by:

$$ \theta^{(t)}= \theta^{(t-1)} + \eta \, \nabla l(\theta^{(t-1)}) $$
where $\eta$ is the step size, and $\nabla l(\theta^{(t-1)} = \left( \dfrac{\partial l(\theta^{(t-1)})}{\partial \beta_{0i}},\dfrac{\partial l(\theta^{(t-1)})}{\partial \beta_{li}}, \dfrac{\partial l(\theta^{(t-1)})}{\partial \gamma_{0j}}, \dfrac{\partial l(\theta^{(t-1)})}{\partial \gamma_{mj}}\right). $ 

The iterations are continued till convergence,the convergence criterion being:

$$ max( \vert{\theta^{(t)} - \theta^{(t-1)}}\vert ) < 0.0001 $$

\section{Stroke Data}\label{study}

We retrospectively analyzed the data of 275 stroke patients admitted to 
a local hospital\footnote{name undisclosed to preserve anonymity in blind submission}.  
Only ischemic stroke patients are considered.
The age group is restricted to 45 -- 75. 
Patients suffering from cancer or severe liver/kidney disease,
patients in critical care and moribund or comatose patients are excluded from the study.

The data includes the variables listed in table \ref{data} for each patient. 
Summary statistics for each of the variables are shown in 
tables in the following sections.
These tables also describe the datatype and the encoding used for categorical variable.
Not all investigations are conducted for all the patients and
the treatment variables differ across patients. Hence the data contains a large number of missing values.
The number of missing values for each variable is also shown in the tables below.


\begin{table}[H]
\footnotesize
\centering
\begin{tabular}{|p{0.75cm}|p{2.1cm}|p{4.5cm}|}
\hline
\multirow{4}{*}{$X_{N \times L}$} & Demographic & Age, Gender, Religion \\
& Time & Dates of event, start of treatment, admission, discharge\\
& Addictions & Smoking, Alcohol\\
& Preconditions & Hypertension, Diabetes, Ischemic Heart Condition, Preceding Fever\\
\hline
$c_I$ & Initial condition & Rankin Score at admission\\
\hline
\multirow{4}{*}{$Y_{N \times M}$} & Investigations & 	Hemoglobin,	Total Counts,	Differential Counts, Platelet, Creatinine, Serum Sodium, Cholesterol, High Density Lipoprotein (HDL), Low Density Lipoprotein (LDL), Triglycerides,
Admission Blood Pressure, Electrocardiogram (ECG), Echo, Doppler \\
& Treatment & Aspirin, Clopidogrel, Atorvastatin, Finofibrate, Edaravon, Citicoline, Heparin, Dalteparin, Enoxaparine, Warfarin, Acitrom, t-PA, , Dabigatran, Anti-hypertensives (ACEI, Beta Channel Blockers, Diuretics, ARB), Other drugs (Piracetam, Mannitol, B-Complex, Pantoprazole, Antibiotics), Physiotherapy\\
& Others & Type of ward, Complication\\
\hline
$c_F$ & Final Condition & Rankin Score at discharge\\
\hline	
\end{tabular}
\caption{Stroke Dataset Variables.}
\label{data}
\end{table}

For our dataset, the contingency table, $\Omega$ is as follows.
The row headers are the values of $c_I$, the initial condition at admission, 
and the column headers are the values of $c_F$, the final condition at discharge,
both measured by the Rankin scale. The $(i,j)^{th}$ entry is the number of patients
with $c_I = i, c_F = j$.

 $$
\kbordermatrix{
    \mbox{Score}&1 &2 &3 &4 &5\\
    1 &22 &0 &0 &0 &0  \\
    2 &40 &29 &1 &0 &0 \\
    3 &1  &44 &25 &0 &2 \\
    4 &0 &0 &38 &34 &2 \\
    5 &0 &0 &2 &15 &20   
}$$

\subsection{Variables in $X_{N \times L}$}

These variables include demographic variables, and variables indicating 
addictions and preconditions. All these are relevant prior to admission
when the initial condition $c_I$ is measured.

\begin{table}[H]
\footnotesize
\centering
\begin{tabular}{|c|c|c|l|}
\hline
Name & Categorical & NA? & Summary Statistics\\
\hline
Age & N & N & Min: 45	Med: 58	Mean: 58.32	Max: 75	SD: 8.61\\
\hline
Gender & Y & N & Male: 169, Female: 106\\
\hline
Religion & Y & N & R1: 209, R2: 41, R3: 25, R4: 4, R5: 5, Others: 6 \\
\hline
\end{tabular}
\caption{Summary Statistics of Demographic Variables. Ri, i $ = 1, \ldots, 5$ are five different religions; Y: Yes, N: No; Min: Minimum, Max: Maximum, SD: Standard Deviation, Med: Median; NA?: Are there missing values?.}
\label{demo}
\end{table}

\begin{table}[H]
\footnotesize
\centering
\begin{tabular}{|c|c|c|c|}
\hline
Name & Categorical & NA? & Count Statistics\\
\hline
Smoking & Y & N & Smokers: 81	Non--smokers: 194\\
\hline
Alcohol & Y & N & Regular: 65	Not regular: 210\\
\hline
\end{tabular}
\caption{Summary Statistics of Addiction Variables. Y: Yes, N: No; NA?: Are there missing values?.}
\label{addic}
\end{table}

\begin{table}[H]
\footnotesize
\centering
\begin{tabular}{|c|c|c|l|}
\hline
Name & Categorical & NA? & Summary Statistics\\
\hline
Hypertension & Y & Y: 3 & Yes: 149	No: 123\\
\hline
Diabetes & Y & Y: 1 &	Yes: 88	No: 186\\
\hline
Ischemic Heart Condition & Y & N & 	Yes: 31	No: 244\\
\hline
Days of preceding fever & N & Y: 13 & Min: 0, Med: 0, Mean: 0.3, Max: 20, SD: 1.95\\
\hline
\end{tabular}
\caption{Summary Statistics of Precondition Variables. Y: Yes, N: No; Min: Minimum, Max: Maximum, SD: Standard Deviation, Med: Median; NA?: Are there missing values, if yes, how many?}
\label{precond}
\end{table}

\subsection{Variables in $Y_{N \times M}$}

These variables include values of various clinical investigations
and treatments given between admission and discharge, i.e.
between the initial condition $c_I$ and final condition $c_F$
measured by the Rankin scale.

\begin{table}[H]
\footnotesize
\centering
\begin{tabular}{|c|c|c|c|c|c|c|c|c|c|}
\hline
Name & Units & NA? & Min	& 1st Q	& Med & Mean	& 3rd Q	& Max & SD\\
\hline
Hemoglobin	&	g/dl & Y:5 &	5.6	& 11.9	& 13.4	& 13.33	& 14.9	& 20.6	& 2.279\\
\hline
Total Counts	&	/ul	& Y:4 & 948	& 7680	& 9440	& 10090	& 11400	& 70000	& 4986.96\\									
\hline
Neutrophils	&	\% & Y:16 &	0	& 62	& 69	& 69.8 & 78.5 &	95	& 12.682\\
\hline
Lymphocytes	&	\% & Y:16 & 0 &	16	& 24 & 23.31 & 	30 & 52.6 &	9.788\\
\hline
Eosinophils	& \% & Y:64 & 0	& 1	& 2	& 3.749	& 4.75 & 61	&	5.764\\
\hline
Platelet count	& lakh/ul & Y:31 & 0 & 1.93	& 2.3 &	3.522 &	2.8	& 261 & 16.582\\
\hline
Creatinine	& mg/dl	& Y:21 & 0.1 & 0.8 & 1 & 1.038 & 1.1 & 7.1 &	0.525\\
\hline
Serum Sodium	& mEq/L	& Y:60 & 118 & 133 & 135 &	135.4 &	137 & 150 & 3.811\\
\hline
Total Cholesterol	& mg/dl	& Y:37 & 1.8 & 143 & 180 & 176.4 & 207 & 379 &	47.284\\
\hline
HDL &	mg/dl	& Y:35 & 3 & 27.75 & 34 & 36.19 & 42 & 182 & 15.562\\
\hline
LDL &	mg/dl	& Y:35 & 22 & 91 & 118 & 119.3 & 141.2 & 902 & 63.144\\
\hline
Triglycerides &	mg/dl & Y:38 & 27 &	85 & 123 & 141 & 180 & 584 & 79.671\\
\hline
Systolic BP & mmHg & Y:10 & 90 & 130 & 150 & 149.2 & 170 & 230 & 26.848\\
\hline
Diastolic BP &	mmHg & Y:10 & 30 & 80 &	90 & 89.34 & 100 & 140 & 14.308\\
\hline
Echo EF & \%	& Y:38 &	0	&60 &	67&64.07	&70&	90	&11.626\\
\hline
\end{tabular}
\caption{Summary Statistics of Investigation Variables. Y: Yes, N: No; Min: Minimum, Med: Median, Max: Maximum, SD: Standard Deviation, 1st Q: first quartile, 3rd Q: third quartile; NA?: Are there missing values, if yes, how many?}
\label{labs}
\end{table}

\begin{table}[H]
\footnotesize
\centering
\begin{tabular}{|c|c|c|c|}
\hline
Name & Normal & Abnormal & NA?\\
\hline
Doppler	&66	&152	&57\\
\hline
ECG	&126&	130&	19\\
\hline
Echo&	53&	194	&28\\
\hline
\end{tabular}
\caption{Summary Statistics of Radiology Variables; encoded as binary based on findings by consultant doctor. NA?: Are there missing values, if yes, how many?}
\label{radio}
\end{table}

\begin{table}[H]
\footnotesize
\centering
\begin{tabular}{|c|c|c|c|}
\hline
Name & Prescribed & Not Prescribed & NA?\\
\hline
Edaravone			&	80	& 	190	&5\\
\hline
Clopidogrel			&	153	&	119	&3\\
\hline
Citicoline			&	33	&	242	&0\\
\hline
Heparin				&	258	&	15	&2\\
\hline
Warfarin			&	218	&	53	&4\\
\hline
t-PA				&	268	&	6	&1\\
\hline
ACEI				&	196	&	79	&0\\
\hline
Beta Channel Blocker&	220	&	55	&0\\
\hline
Diuretics			&	240	&	26	&9\\
\hline
ARB					&	238	&	30	&7\\
\hline
Piracetam			&	238	&	34	&3\\
\hline
B. Complex			&	191	&	84	&0\\
\hline
Met/Glyco			&	197	&	76	&2\\
\hline
Insulin				&	196	&	79	&0\\
\hline
Pantoprazole		&	145	&	129	&1\\
\hline
Physiotherapy		&	200	&	73	&2\\
\hline
Antibiotics			&	197	&	76	&2\\
\hline
Finofibrate			&	264	&	3	&8\\
\hline
Acitrom				&	270	&	5	&0\\
\hline
\end{tabular}
\caption{Summary Statistics of Treatment Variables; encoded as binary based on prescriptions of consultant doctor. NA?: Are there missing values, if yes, how many?}
\label{treatment}
\end{table}

\begin{table}[h]
\footnotesize
\centering
\begin{tabular}{|c|c|c|c|c|c|c|c|c|c|}
\hline
Name & Units & NA? & Min	& 1st Q	& Med & Mean	& 3rd Q	& Max & SD\\
\hline
Aspirin &	mg&	N&	1	&2	&2	&3.105	&5	&10	&1.771\\
\hline
Atorvastatin&	mg	&N	&0	&40	&40	&37.98	&40	&80	&12.915\\
\hline
Dalteparine	&units	&N	&0	&0	&0	&1973	&5000	&5000	&2188.849\\
\hline
Enoxaparine	&ml	&N	&0	&0	&0	&0.04364	&0	&0.6	&0.144\\
\hline
\end{tabular}
\caption{Summary Statistics of Treatment Variables where dosages were used in data; 0 indicates not prescribed and nonzero value is the dosage prescribed. Y: Yes, N: No; Min: Minimum, Med: Median, Max: Maximum, SD: Standard Deviation, 1st Q: first quartile, 3rd Q: third quartile; NA?: Are there missing values, if yes, how many?}
\label{dosage}
\end{table}

\subsection{Univariate Analysis}

\subsubsection{ANOVA}
Univariate One Way Analysis of Variance(ANOVA) is performed on each predictor variable with the Rankin Score at Discharge denoting the groups. The test shows how each of the variables varies between the groups(i.e high between-group variance and low within-group variance). Table \ref{anova} shows the p-values corresponding to testing of hypothesis $H_{0}$ : the variable does not vary across groups , against $H_{1}$ : The variable varies across the groups.

\begin{table}[h]
\footnotesize
\centering
\begin{tabular}{|c|c|c|}
\hline
Variables & p-value & Status\\
\hline
Stay in hospital(Days)	&0&	Reject\\
\hline
NEUTROPHILS & 0	& Reject\\
\hline
LYMPHOCYTES	&0	&Reject\\
\hline
Piracetam	&0	&Reject\\
\hline
Physiotherapy	&0&	Reject\\
\hline
Complication	&0&	Reject\\
\hline
Rankin Score at Admission	&0&	Reject\\
\hline
Antibiotics	&0.001&	Reject\\
\hline
Heparin	&0.002&	Reject\\
\hline
TC	&0.005&	Reject\\
\hline
Doppler	&0.009&	Reject\\
\hline
Platelet	&0.019	&Reject\\
\hline
Cholesterol	&0.02&	Reject\\
\hline
LDL	&0.025&	Reject\\
\hline
Diastolic	&0.038&	Reject\\
\hline
Systolic	&0.045	&Reject\\
\hline
Warfarin	&0.048&	Reject\\
\hline
Time between event and Treatment(Days)	&0.075&	Accept\\
\hline
Serum Na.	&0.115&	Accept\\
\hline
Dalteparine	&0.116&	Accept\\
\hline
Ward	&0.128&	Accept\\
\hline
Aspirin	&0.164&	Accept\\
\hline
Pantoprazole	&0.165&	Accept\\
\hline
Preceding fever	&0.216&	Accept\\
\hline
EOSINOPHILS	&0.241&	Accept\\
\hline
Hypertension	&0.249&	Accept\\
\hline
Creatinine	&0.283&	Accept\\
\hline
Beta Channel Blocker	&0.314&	Accept\\
\hline
Religion	&0.317&	Accept\\
\hline
Citicoline	&0.325&	Accept\\
\hline
Alcohol abuse	&0.337&	Accept\\
\hline
Atorvastatin	&0.357&	Accept\\
\hline
Enoxaparine	&0.386&	Accept\\
\hline
Acitrom	&0.394&	Accept\\
\hline
ARB	&0.424&	Accept\\
\hline
Gender	&0.428&	Accept\\
\hline
EF	&0.431&	Accept\\
\hline
B Complex	&0.457&	Accept\\
\hline
Isch Heart Condition	&0.548&	Accept\\
\hline
Edaravone	&0.561&	Accept\\
\hline
Clopidogrel	&0.567&	Accept\\
\hline
Met/Glyco	&0.57&	Accept\\
\hline
Smoking	&0.571&	Accept\\
\hline
Triglycerides	&0.573&	Accept\\
\hline
Echo	&0.578&	Accept\\
\hline
Insulin	&0.587&	Accept\\
\hline
Diabetes	&0.591&	Accept\\
\hline
HDL	&0.618&	Accept\\
\hline
ECG	&0.691&	Accept\\
\hline
Finofibrate	&0.708&	Accept\\
\hline
ACEI	&0.736&	Accept\\
\hline
Age	&0.867&	Accept\\
\hline
Diuretics	&0.888&	Accept\\
\hline
t.PA	&0.915&	Accept\\
\hline
Hb	&0.922&	Accept\\
\hline
\end{tabular}
\caption{P values from ANOVA}
\label{anova}
\end{table}
 
The variables are arranged in increasing order of p-value ; the null hypothesis is accepted with 5\% level of significance if p-value \textgreater \ 0.05.

\subsubsection{Univariate Logistic Regression}

Univariate Logistic Regression is performed on each variable by taking the Rankin Score at Discharge as the response and the variable as the predictor. Table \ref{uniLR} lists the residual deviances for each variable in increasing order (lower the value, higher is the association between and the predictor and response). The lowest value of 446.57 for Rankin score at admission shows the high correlation with outcome at discharge.

\begin{table}[h]
\footnotesize
\centering
\begin{tabular}{|c|c|}
\hline
Variables	&Residual Deviance\\
\hline
Rankin Score at Admission	&446.57\\
\hline
Physiotherapy	&852.09\\
\hline
LYMPHOCYTES	&862.52\\
\hline
Stay in hospital (Days)	&863.60\\
\hline
Complication	&863.77\\
\hline
NEUTROPHILS	&867.11\\
\hline
Piracetam	&868.38\\
\hline
Antibiotics	&871.02\\
\hline
Cholesterol	&874.58\\
\hline
Gender	&874.89\\
\hline
Heparin	&875.16\\
\hline
Diuretics	&875.53\\
\hline
LDL	&875.73\\
\hline
TC	&876.00\\
\hline
Diastolic	&876.06\\
\hline
Doppler    	&876.36\\
\hline
Diabetes	&878.26\\
\hline
Platelet	&879.01\\
\hline
Time between event and treatment (Days)	&879.22\\
\hline
Systolic	&879.29\\
\hline
EOSINOPHILS	&879.45\\
\hline
Serum Na.	&880.12\\
\hline
ARB	&880.29\\
\hline
Insulin	&880.32\\
\hline
Smoking	&881.13\\
\hline
Religion	&881.14\\
\hline
Finofibrate	&881.37\\
\hline
Isch. Heart Condition	&882.06\\
\hline
HDL	&882.10\\
\hline
Warfarin	&882.57\\
\hline
Hb	&882.66\\
\hline
Pantoprazole	&882.73\\
\hline
Acitrom	&883.16\\
\hline
Dalteparine	&883.25\\
\hline
Aspirin	&883.85\\
\hline
Ward	&883.97\\
\hline
Met/Glyco	&884.27\\
\hline
Creatinine	&884.28\\
\hline
Clopidogrel	&884.38\\
\hline
Triglycerides	&884.40\\
\hline
Citicoline	&884.52\\
\hline
Enoxaparine	&884.69\\
\hline
EF	&884.82\\
\hline
Atorvastatin	&884.98\\
\hline
Alcohol.abuse	&885.06\\
\hline
ACEI	&885.29\\
\hline
t.PA	&885.81\\
\hline
Preceding fever	&885.87\\
\hline
Age	&886.07\\
\hline
Edaravone	&886.66\\
\hline
Beta Channel Blocker	&886.80\\
\hline
Hypertension	&887.04\\
\hline
Echo	&887.14\\
\hline
B. Complex	&887.28\\
\hline
ECG	&887.31\\
\hline

\end{tabular}
\caption{Residual Deviance from Univariate Logistic Regression}
\label{uniLR}
\end{table}

\clearpage

\section{Simulated Data}

We test our model on 4 sets of synthetic data where the contingency tables are
simulated from our dataset by multiplying each element of the reduced contingency matrix from the stroke dataset by 2,10,15 and 20.
$$
\kbordermatrix{
    \mbox{Score}&g_1 &g_2 &g_3\\
    g_1&22 &0 &0  \\
    g_2&41 &99 &2 \\
    g_3&0 &40 &71  
}$$

We generate $x_k$  and $y_k$ by simulating respectively from  6-dimensional normal distributions $\mathcal{N}(\mu _{Xi},I)$ and  $\mathcal{N}(\mu _{Yj},I)$; where $ \mu _{Xi} = \begin{pmatrix}
																	  i\\
																	  i+1\\
																	  i+2\\
																	  i+3\\
																	  i+4\\
																	  i+5
																	  \end{pmatrix}$ 
, $ \mu _{Yj} = \begin{pmatrix}
			4-j\\
			5-j\\
			6-j\\
		    7-j\\
			8-j\\
			9-j\\
			\end{pmatrix}$ 		
and $I$ denotes the identity matrix. $i$ and $j$ respectively are the values of $c_I$ (row) and $c_F$ (column).

\begin{itemize}

\item\textbf{{Simulation 1}}

\textbf{Contingency Table :}

$$
\kbordermatrix{
    \mbox{Score}&g_1 &g_2 &g_3\\
    g_1&44 &0 &0  \\
    g_2&82 &198 &4 \\
    g_3&0 &80 &142
}$$

\textbf{Results:}

\begin{table}[htb]
\centering
\begin{tabular}{|c|c|c|}
\hline
Method & Mean Accuracy & SD \\
\hline
$LR_{row}$ & 72.3\% & 2.45 \\
\hline
$SVM_{row}$ & 71.5\% & 2.89 \\
\hline
$LR$ &73.1\% & 2.54\\
\hline
$SVM$ & 73.5\% & 2.27  \\
\hline
NEW & \textbf{77.9\%} & 3.1\\
\hline
\end{tabular}
\caption{Average accuracy and standard deviation (SD) over five--fold cross validation on simulated data 1}
\label{sim1}
\end{table}

\item\textbf{{Simulation 2}}

\textbf{Contingency Table :}

$$
\kbordermatrix{
    \mbox{Score}&g_1 &g_2 &g_3\\
    g_1&220 &0 &0  \\
    g_2&410 &990 &20 \\
    g_3&0 &400 &710
}$$

\textbf{Results:}

\begin{table}[htb]
\centering
\begin{tabular}{|c|c|c|}
\hline
Method & Mean Accuracy & SD \\
\hline
$LR_{row}$ & 77.4\% & 2.15 \\
\hline
$SVM_{row}$ & 77.7\% & 2.66 \\
\hline
$LR$ & 79\% & 2.46\\
\hline
$SVM$ & 78.8\% &  2.80\\
\hline
NEW & \textbf{84.3\%} & 2.99 \\
\hline
\end{tabular}
\caption{Average accuracy and standard deviation (SD) over five--fold cross validation on simulated data 2}
\label{sim2}
\end{table}

\item\textbf{{Simulation 3}}

\textbf{Contingency Table :}

$$
\kbordermatrix{
    \mbox{Score}&g_1 &g_2 &g_3\\
    g_1&330 &0 &0  \\
    g_2&615 &1485 &30 \\
    g_3&0 &600 &1065
}$$

\textbf{Results:}

\begin{table}[htb]
\centering
\begin{tabular}{|c|c|c|}
\hline
Method & Mean Accuracy & SD \\
\hline
$LR_{row}$ & 79.8\% & 1.97 \\
\hline
$SVM_{row}$ & 80.6\% & 2.16 \\
\hline
$LR$ & 82.4\% & 2.38 \\
\hline
$SVM$ & 81.7\% & 2.84 \\
\hline
NEW & \textbf{91.2\%} & 3.26 \\
\hline
\end{tabular}
\caption{Average accuracy and standard deviation (SD) over five--fold cross validation on simulated data 3}
\label{sim3}
\end{table}

\item\textbf{{Simulation 4}}

\textbf{Contingency Table :}

$$
\kbordermatrix{
    \mbox{Score}&g_1 &g_2 &g_3\\
    g_1&440 &0 &0  \\
    g_2&820 &1980 &40 \\
    g_3&0 &800 &1420
}$$

\textbf{Results:}

\begin{table}[htb]
\centering
\begin{tabular}{|c|c|c|}
\hline
Method & Mean Accuracy & SD \\
\hline
$LR_{row}$ & 83.0\% & 2.11 \\
\hline
$SVM_{row}$ & 82.4\% &  1.70\\
\hline
$LR$ & 85.1\%  & 2.54 \\
\hline
$SVM$ & 85.2\% & 2.41 \\
\hline
NEW & \textbf{92.5\%} & 2.45  \\
\hline
\end{tabular}
\caption{Average accuracy and standard deviation (SD) over five--fold cross validation on simulated data 4}
\label{sim4}
\end{table}

\end{itemize}

\subsubsection{Simulation for varying K with N=5500}

\begin{itemize}

\item\textbf{$K=3$}

\textbf{Contingency Table :}

$$
\kbordermatrix{
    \mbox{Score}&g_1 &g_2 &g_3\\
    g_1&440 &0 &0  \\
    g_2&820 &1980 &40 \\
    g_3&0 &800 &1420
}$$

\textbf{Results :}

\begin{table}[htb]
\centering
\begin{tabular}{|c|c|c|}
\hline
Method & Mean Accuracy & SD \\
\hline
$LR_{row}$ & 83.0\% & 2.11 \\
\hline
$SVM_{row}$ & 82.4\% &  1.70\\
\hline
$LR$ & 85.1\%  & 2.54 \\
\hline
$SVM$ & 85.2\% & 2.41 \\
\hline
NEW & \textbf{92.5\%} & 2.45  \\
\hline
\end{tabular}
\caption{Average accuracy and standard deviation (SD) over five--fold cross validation on simulated data for K=3}
\end{table}

\item\textbf{$K=4$}

\textbf{Contingency Table :}

$$
\kbordermatrix{
    \mbox{Score}&g_1 &g_2 &g_3 &g4\\
    g_1&270 &170 &0 &0  \\
    g_2&300 &990 &90 &40 \\
    g_3&230 &300 &790 &100\\	
    g_4&0 &300 &720 &1200
}$$

\textbf{Results :}

\begin{table}[htb]
\centering
\begin{tabular}{|c|c|c|}
\hline
Method & Mean Accuracy & SD \\  
\hline
$LR_{row}$ & 79.8\% & 2.8 \\
\hline
$SVM_{row}$ & 80.1\% &  1.94\\
\hline
$LR$ & 85\%  & 2.67 \\
\hline
$SVM$ & 82.8\% & 2.34 \\
\hline
NEW & \textbf{90.02\%} & 2.78  \\
\hline
\end{tabular}
\caption{Average accuracy and standard deviation (SD) over five--fold cross validation on simulated data for K=4}
\end{table}

\item\textbf{$K=5$}

\textbf{Contingency Table :}

$$
\kbordermatrix{
    \mbox{Score}&g_1 &g_2 &g_3 &g4 &g_5\\
    g_1&270 &110 &60 &0 &0  \\
    g_2&100 &580 &110 &50 &20 \\
    g_3&80	&200 &420 &50 &10\\
    g_4&70 &150 &260 &740 & 200\\	
    g_5&0 &200 &320 &500 &1000
}$$

\textbf{Results :}

\begin{table}[htb]
\centering
\begin{tabular}{|c|c|c|}
\hline
Method & Mean Accuracy & SD \\  
\hline
$LR_{row}$ & 77.9\% & 2.67 \\
\hline
$SVM_{row}$ & 77.5\% &  2.36\\
\hline
$LR$ & 83.8\%  & 2.76 \\
\hline
$SVM$ & 80.6\% & 2.72 \\
\hline
NEW & \textbf{87.2\%} & 3.23  \\
\hline
\end{tabular}
\caption{Average accuracy and standard deviation (SD) over five--fold cross validation on simulated data for K=5}
\end{table}

\item\textbf{$K=6$}

\textbf{Contingency Table :}

$$
\kbordermatrix{
    \mbox{Score}&g_1 &g_2 &g_3 &g4 &g_5 &g6\\
    g_1&300 &100 &70 &30 &0 &0 \\
    g_2&120 &480 &150 &80 &20 &10 \\
    g_3&80	&150 &370 &100 &50 &10\\
    g_4&70 &150 &200 &740 & 160 &100\\	
    g_5&30 &70 &110 &200 &500 &100\\
    g_6&0 &25 &60 &150 &225 &670\\
}$$

\textbf{Results :}

\begin{table}[htb]
\centering
\begin{tabular}{|c|c|c|}
\hline
Method & Mean Accuracy & SD \\  
\hline
$LR_{row}$ & 74.3\% & 2.83\\
\hline
$SVM_{row}$ & 73\% &  2.7\\
\hline
$LR$ & 79.9\%  & 2.99 \\
\hline
$SVM$ & 76.2\% & 3.21 \\
\hline
NEW & \textbf{85.9\%} & 3.10 \\
\hline
\end{tabular}
\caption{Average accuracy and standard deviation (SD) over five--fold cross validation on simulated data for K=6}
\end{table}

\end{itemize}

\end{document}